\documentclass[10pt,prb,onecolumn, superscriptaddress, footinbib]{revtex4}
\usepackage[pdftex]{graphicx}
 \usepackage{amsmath}
\usepackage[T1]{fontenc}
\usepackage{flushend}
\usepackage{amssymb}
\usepackage{amsfonts}
\usepackage{bm}
 \usepackage{amsmath} 
\usepackage{lipsum}
\usepackage{amsfonts} 
\usepackage{amssymb, mathrsfs}
\usepackage{braket}
\usepackage{graphicx} 
\usepackage{subfigure}
\usepackage{bbm}

\def\beq{\begin{equation}}
\def\eeq{\end{equation}}
\def\bsp{\begin{split}}
\def\esp{\end{split}}
\def\bea{\begin{eqnarray}}
\def\eea{\end{eqnarray}}
\def\ba{\begin{array}}
\def\ea{\end{array}}

\def\dg{\dagger}

\def\lb{\left(}
\def\rb{\right)}

\def\l.{\left.}
\def\r.{\right.}

\def\ra{\rangle}
\def\la{\langle}

\def\bo{{\vec k}}

\begin{document}

\date{\today}
\title{\large Floquet Weyl Magnons in Three-Dimensional Quantum Magnets}
\email{sowerre@perimeterinstitute.ca}
\author{S. A. Owerre}
\affiliation{Perimeter Institute for Theoretical Physics, 31 Caroline St. N., Waterloo, Ontario N2L 2Y5, Canada.}

\maketitle

\noindent\textbf {
In three-dimensional (3D) quantum magnets, magnonic Weyl points (WPs) featuring linear band crossing of two non-degenerate magnon branches can emerge in certain lattice geometry when time-reversal symmetry is broken macroscopically. Unfortunately, there are very limited 3D quantum magnets that host magnonic WPs, and they are yet to be observed experimentally because the intrinsic perturbative interactions that break time-reversal symmetry macroscopically can be very negligible.  Here, we present an alternative means via photo-irradiation, in which magnonic WPs can emerge in 3D quantum magnets without relying on intrinsic perturbative interactions to break time-reversal symmetry. By utilizing the magnonic Floquet-Bloch theory, we put forward the general theory of magnonic Floquet WPs in 3D quantum magnets. We show that periodically driven 3D magnonic Dirac nodal-line (DNL) and 3D magnonic gapped trivial insulators can generate  3D magnonic Floquet WPs, which can be tuned by the incident circularly-polarized light. We demonstrate the existence of magnonic Floquet WPs by combining the study of the magnon dispersions, Berry curvatures, and the anomalous thermal Hall effect. The general theoretical formalism  can be applied to different magnetic insulators, and thus extending the concept of magnonic WPs to a broader class of 3D magnetically ordered systems.  
}

\vspace{10px}

The condensed matter realization of Weyl semimetals  as emergent quasiparticles hosting Weyl fermions has attracted considerable interest in recent years \cite{wan,bur,xu,lv}. Weyl semimetals are considered to be the first material realization of Weyl fermions in nature. Generically, WPs are allowed in 3D solid-state materials with either broken  inversion ($\mathcal P$) or time-reversal ($\mathcal T$) symmetry. This guarantees that two WPs separated in momentum space are topologically stable and can only disappear by pair annihilation \cite{vol,vol1}.  Essentially, the general notion of WPs in condensed-matter systems is manifested when two  non-degenerate topologically protected bands  cross linearly in 3D momentum space. This linear band crossing point is independent of the quasiparticle excitations and their corresponding  quantum statistics.  Therefore, it occurs in both bosonic and fermionic systems.  Recently, magnonic WPs  have come into focus \cite{mw1, mw2, mw3, mw4, mw5, mw6, mw7, mw8} as the bosonic analogs of electronic WPs, and  they occur in 3D (as well as quasi-2D) insulating ordered magnets when two non-degenerate magnon branches cross linearly in the 3D Brillouin zone (BZ).
 
In magnetic Weyl systems,   $\mathcal T$-symmetry is naturally broken owing to  the presence of magnetic order. Nonetheless, magnonic WPs generally do not exist in every 3D magnetic material. The existence of stable magnonic WPs can be achieved when  $\mathcal T$-symmetry is macroscopically (explicitly) broken. For insulating quantum ferromagnets, macroscopically broken $\mathcal T$-symmetry can be achieved  by the combination of spontaneous magnetization and the Dzyaloshinskii-Moriya (DM) interaction \cite{dm,dm2} in the direction of the magnetization.  The DM interaction is allowed in quantum magnets that lack an inversion center, and it plays the role of spin-orbit coupling (SOC).  For insulating quantum antiferromagnets, however,  the antiferromagnetic order can be  restored by  symmetry, hence  the spontaneous magnetization and the DM interaction can be inadequate to provide stable magnonic WPs in antiferromagnets. In this case, one can only achieve magnonic WPs through symmetry-breaking noncoplanar spin textures with nonzero scalar spin chirality or applied external magnetic field. The former provides a possible transition to chiral spin liquids in which $\mathcal T$-symmetry is also broken macroscopically.  

One of the hallmarks of Weyl semimetals is the appearance of the Fermi arc surface states, which connect the surface projection of WPs  in momentum space  \cite{wan,bur}. This provides a distinct topological classification from topological insulators (TIs). Besides,  WPs are also sinks and sources of the Berry curvature. In other words, a single WP acts as a (magnetic) monopole of the Berry curvature in momentum space.  Similarly,    magnonic WPs also host magnon arc surface states as  a topological feature, and they  are also the  monopoles of the Berry curvature  in momentum space.  Despite the simplicity of the theoretical concepts of WPs in quantum materials, their experimental realizations  in real materials  are elusive. This is in part due to the fact  that  the intrinsic perturbative interactions  that are necessary for WPs to occur can be very weak or the quantum materials may have strong correlated many-body effects. Thus far, the experimental realizations of bosonic WPs have only been reported in artificial photonic and phononic optical systems \cite{lu,fee}. Therefore, it is desirable to explore other possibilities in which bosonic magnetic WPs can be witnessed in quantum materials.  
 
In recent years, photo-irradiation  of solid-state materials  have emerged as an alternative means to extend the search for topological quantum materials \cite{abso}. By exposing a topologically trivial quantum material to a time-periodic electromagnetic (laser) field, the  intrinsic properties of the material can be altered via light-matter interactions. Basically, the charge carriers in the quantum material couple to the time-periodic vector potential through a time-dependent Peierls phase, in a similar way to the Aharonov-Bohm phase \cite{aha1}. Consequently,  the quantum material becomes a periodically driven system, which can be studied by the Floquet-Bloch theory. The resulting effect of irradiated quantum materials is that $\mathcal T$-symmetry breaking terms can be photo-induced, leading  to different nontrivial Floquet topological phases such as  Floquet topological insulators \cite{tp1,tp2,tp2a,tp3,tp4,tp4a,tp5,tp6,tp7,tp8,tp9,tp10, tp6a, roy,roy1} and Floquet Weyl semimetals \cite{we1,we2,we3,we4,we5,we6, we7, we7a, we4a}. 
   
 In fact,  the mechanism of  photo-irradiation is not restricted to electronic charge materials, but also applies  to solid-state materials with charge-neutral carriers.  In particular, charge-neutral magnons are simply  magnetic dipole moments hopping in an ordered magnetic insulator, and they produce a force similar to the Lorentz force on charged particles \cite{loss}. Therefore, magnons can also  couple to a time-independent electric  field through the Aharonov-Casher effect \cite{aha, ahat, ahaz} --- a mechanism in which charge-neutral particles acquire a geometric phase in an electric field background. In this formalism,   magnonic Landau levels can be induced  in insulating magnets \cite{mei}, and chiral anomaly can be induced in Weyl magnons \cite{mw3, mw5}, in analogy to  electronic systems. However, the physics of time-independent electric field is completely different from that of time-periodic electric field from a laser source. In the latter,  one realizes  a time-dependent version of the Aharonov-Casher effect (see Methods), which leads to periodically driven magnetic insulators also amenable to solution via the Floquet-Bloch theory. The magnonic Floquet-Bloch theory describes the interaction of light with magnonic Bloch states in insulating quantum magnets. Consequently, two-dimensional (2D) Dirac magnons in honeycomb ferromagnets can be driven to  2D magnonic Floquet TIs  via a photoinduced next-nearest-neighbour DM interaction \cite{owe}, and also topological phase transition can be photoinduced in intrinsic magnon TIs such as Cu(1-3, bdc) \cite{owe1}. Thus,  magnonic systems can now be studied in analogy to photo-irradiated graphene  and Chern insulators,  which generate 2D electronic Floquet TIs  \cite{tp1,tp4,tp2a}  and photoinduced topological phase transition \cite{tp2} respectively.  

In this report, we generalize this new concept to 3D insulating quantum magnets. In this case the incident light can be applied in different directions due to the 3D nature of the system, but not all directions generate WPs \cite{foot}. We will start with a 3D quantum magnets with Dirac nodal-line (DNL) phase in which the Dirac points (DPs) form closed loops in the BZ. By fine-tuning the model parameters the DNLs can be gapped out to a trivial insulator. Therefore, our 3D quantum magnet has two phases that are topologically trivial.  Our main goal is to generate topologically nontrivial phase from this system by applying photo-irradiation in different directions of the 3D quantum magnet. In particular,  we show that while photo-irradiation in all other direction generates 3D magnonic Floquet TIs, photo-irradiation in the direction parallel to the DNLs  generates 3D magnonic Floquet WPs, which is very similar to electronic Floquet systems \cite{we4a, we4}. We also observe that tunable 3D magnonic Floquet WPs can emerge from periodically driven 3D magnonic gapped trivial insulator  using circularly-polarized lights. We establish a compelling evidence of magnonic Floquet WPs in this 3D insulating quantum magnet by computing the monopole distributions  of the Berry curvature in momentum space and  the thermal Hall conductivity, both of which vanish in quantum magnets with $\mathcal {T}$-symmetry, such as the undriven  Dirac magnons, or DNL magnons, or trivial magnon insulators.  The theoretical formalism and the results are general, and can be applied to different magnetic insulators, including the recently observed Dirac magnons in 3D antiferromagnet Cu$_3$TeO$_6$ \cite{kli, yao,bao}. We envision that our results will greatly impact future research in magnonic topological systems, and extend the experimental search for magnonic WPs to a broader class of 3D quantum magnetic insulators, with potential practical applications to features such as  photo-magnonics \cite{benj}, magnon spintronics \cite{magn, benja},  and ultrafast optical control of magnetic spin currents \cite{ment, tak4, tak4a,walo}.

\vspace{10px}
\noindent \textbf{\large Results}

\begin{figure}
\centering
\includegraphics[width=.8\linewidth]{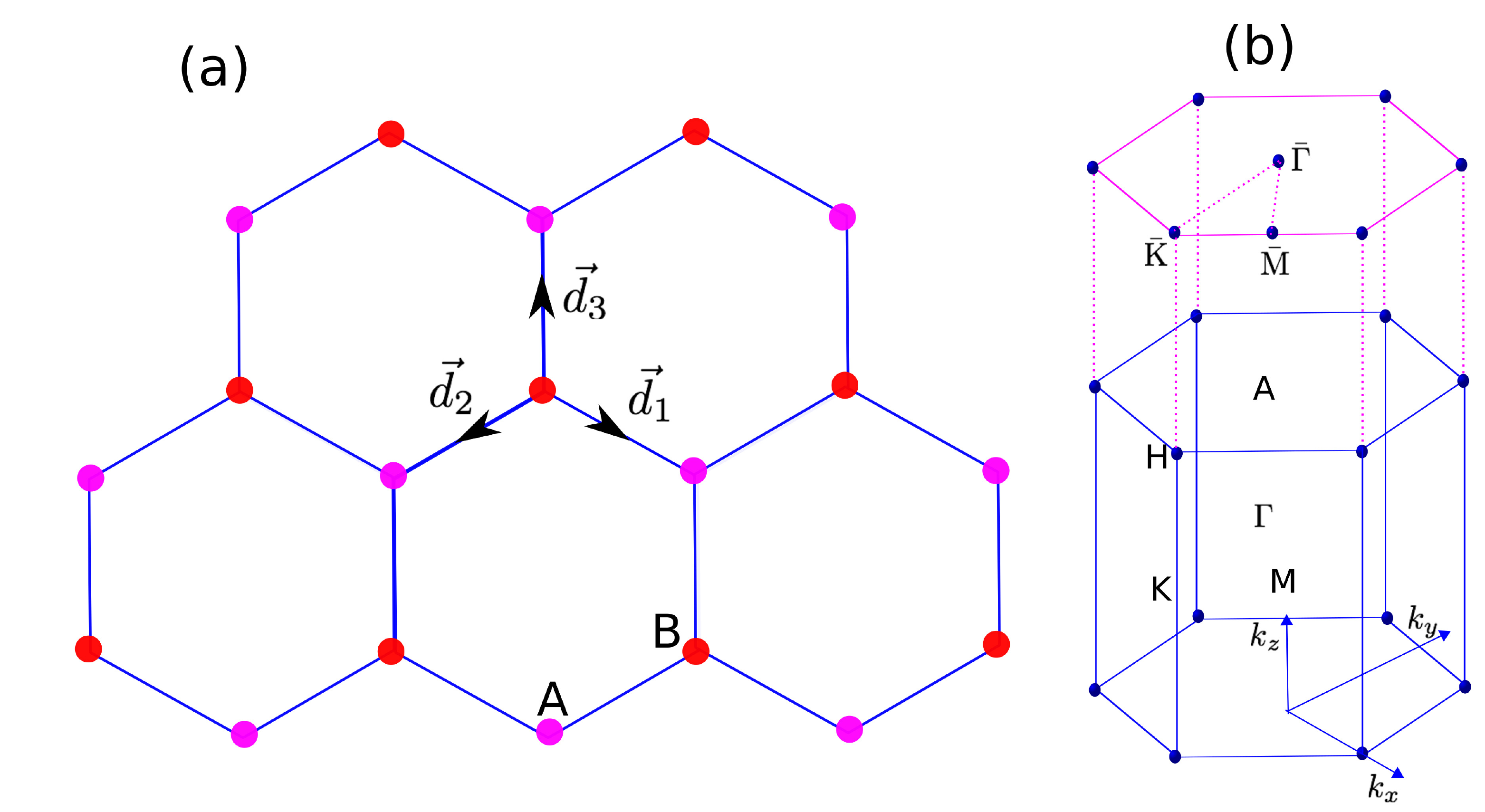}
\caption{Color online. (a) Top view of the honeycomb ferromagnetic layers with vertical bond stacked along the (001) direction.   (b) The bulk Brillouin zone (BZ)  and  its projection onto the hexagonal (001) surface BZ. }
\label{lattice}
\end{figure}

\noindent\textbf{Spin Model.}~~ We study the simple Heisenberg spin Hamiltonian of layered ferromagnets, governed by
\begin{align}
\mathcal H&=-J\sum_{ \la i j\ra,\ell}{\vec S}_{i,\ell}\cdot{\vec S}_{j, \ell}-J_{L}\sum_{\la\ell\ell^\prime\ra, i} {\vec S}_{i,\ell}\cdot{\vec S}_{i,\ell^\prime},
\label{model}
\end{align}
where ${\vec S}_{\ell}=(S^x_{\ell}, S^y_{\ell}, S^z_{\ell})$ is the spin vector at site  ${\ell}$. Here $J$ and $J_{L}$ are  the intralayer and interlayer (vertical bond) ferromagnetic interactions respectively.  The Hamiltonian in Eq.~\ref{model} is applicable to different layered ferromagnets in various lattice geometries. In this report, we will focus on honeycomb layered ferromagnets. In   Fig.~\ref{lattice}(a) and Fig.~\ref{lattice}(b) we have shown the top view of the  honeycomb lattice  stacked with a vertical bond along the (001) direction and its 3D Brillouin zone (BZ) respectively. Indeed, most realistic bulk layered honeycomb ferromagnetic materials such as the honeycomb chromium  compounds CrX$_3$ (X $\equiv$ Br, Cl, and I) \cite{dav0,dav1,dav2,dav3,foot1}, have an inversion center. Therefore, the DM interaction is forbidden by symmetry  in these materials. We would like to mention that the realistic parameter regime of the spin Hamiltonian in Eq.~\eqref{model} is not the main focus in this report.  Our  main objective is to demonstrate how magnonic Floquet WPs can be generated by periodic driving of 3D DNL magnons and 3D trivial magnon insulators, which are obtainable from  Eq.~\eqref{model} in different parameter regimes. In order to achieve this goal, we consider honeycomb ferromagnetic layers stacked similarly to ABC-stacked graphene \cite{ss1,ss2, ss3,ss,ss2a}.

 \vspace{10px} 

\noindent\textbf{Undriven magnonic Dirac nodal-line.}~~ The concept of DNLs emerges when the DPs form a loop or ring in the BZ. This usually happens in 3D systems without explicit $\mathcal T$-symmetry breaking terms or other forms of symmetry protection.  
In this section, we will introduce this concept using the underlying  magnetic excitations of the spin Hamiltonian in Eq.\eqref{model}.  In the  low temperature regime, the magnetic excitations of ordered ferromagnetic insulators are charge-neutral magnons, and they can be captured by the Holstein Primakoff (HP)  transformation \cite{hp}:  $S_{i,\ell}^{ z}= S-a_{i,\ell}^\dagger a_{i,\ell},~S_{i,\ell}^+\approx \sqrt{2S}a_{i,\ell}=(S_{i,\ell}^-)^\dg,$ where $a_{i,\ell}^\dagger (a_{i,\ell})$ are the bosonic creation (annihilation) operators, and  $S^\pm_{i,\ell}= S^x_{i,\ell} \pm i S^y_{i,\ell}$ denote the spin raising and lowering  operators.  The corresponding  non-interacting magnon Hamiltonian is given by $\mathcal H=\sum_{\vec{k}} \psi_{\vec{k}}^\dg \mathcal H({\vec{k}})\psi_{\vec{k}}$ with $\psi_{\vec{k}}^\dg=\big(a_{{\vec k},A}^\dg,a_{{\vec k},B}^\dg\big)$, 
\begin{align}
\mathcal H(\vec k)&=\rho_0{\bf 1}_{2\times 2}+
\begin{pmatrix}
0& \rho(\vec k)\\
\rho^*(\vec k)&0\\
\end{pmatrix},
\label{ham1}
\end{align}
where ${\bf 1}_{2\times 2}$ is an identity matrix.  $\rho_0=3JS + J_LS$ and $\rho(\vec k)=\rho(\vec k_\parallel)+\rho(k_z)$, with  $\rho(k_z)=-t_L\exp(ik_z)$, $\rho(\vec k_{\parallel})=-t\sum_j e^{i\vec{k}_{\parallel}\cdot\vec{d}_j}$. Here,  $t_L=J_LS,~t=JS$,  $\vec d_1=(\sqrt{3}/2, -1/2)$, $\vec d_2=-(\sqrt{3}/2, 1/2)$, and $\vec d_3=(0, 1)$. The total momentum vector is defined as $\vec k =(\vec k_\parallel, k_z)$, where the in-plane wave vector is $ \vec k_{\parallel}=(k_x,k_y)$. Using the Pauli matrices $\sigma_i~(i=x,y,z)$, we write the Hamiltonian \eqref{ham1} as
\begin{align}
\mathcal H(\vec k)=f_0\sigma_0 +f_x(\vec k)\sigma_x+f_y(\vec k)\sigma_y,
\label{nl}
\end{align}
where $\sigma_0\equiv {\bf 1}_{2\times 2}$ and $f_0=\rho_0$,
\begin{align}
&f_x(\vec k)=-t\sum_j\cos(\vec k_\parallel\cdot \vec d_j)-t_L\cos(k_z),\\&
f_y(\vec k)=t\sum_j\sin(\vec k_\parallel\cdot \vec d_j)+t_L\sin(k_z).
\end{align}

 The pseudospin time-reversal symmetry operator is $\mathcal T=\sigma_0\mathcal K$, where $\mathcal K$ is complex conjugation. Evidently,  the Hamiltonian in Eq.~\eqref{nl} is $\mathcal T$-invariant. The condition for DNLs to exist  requires $f_x(\vec k)=f_y(\vec k)=0$. This condition is satisfied in the $k_z=\pi$ plane  at  $k_y=0$ and $k_x=\pm k_x^\text{D}$, where 
\begin{align}
k_x^\text{D}&=\frac{2}{\sqrt{3}}\arccos \lb \frac{-1+t_L/t}{2}\rb.
\end{align}

\begin{figure}
\centering
\includegraphics[width=.8\linewidth]{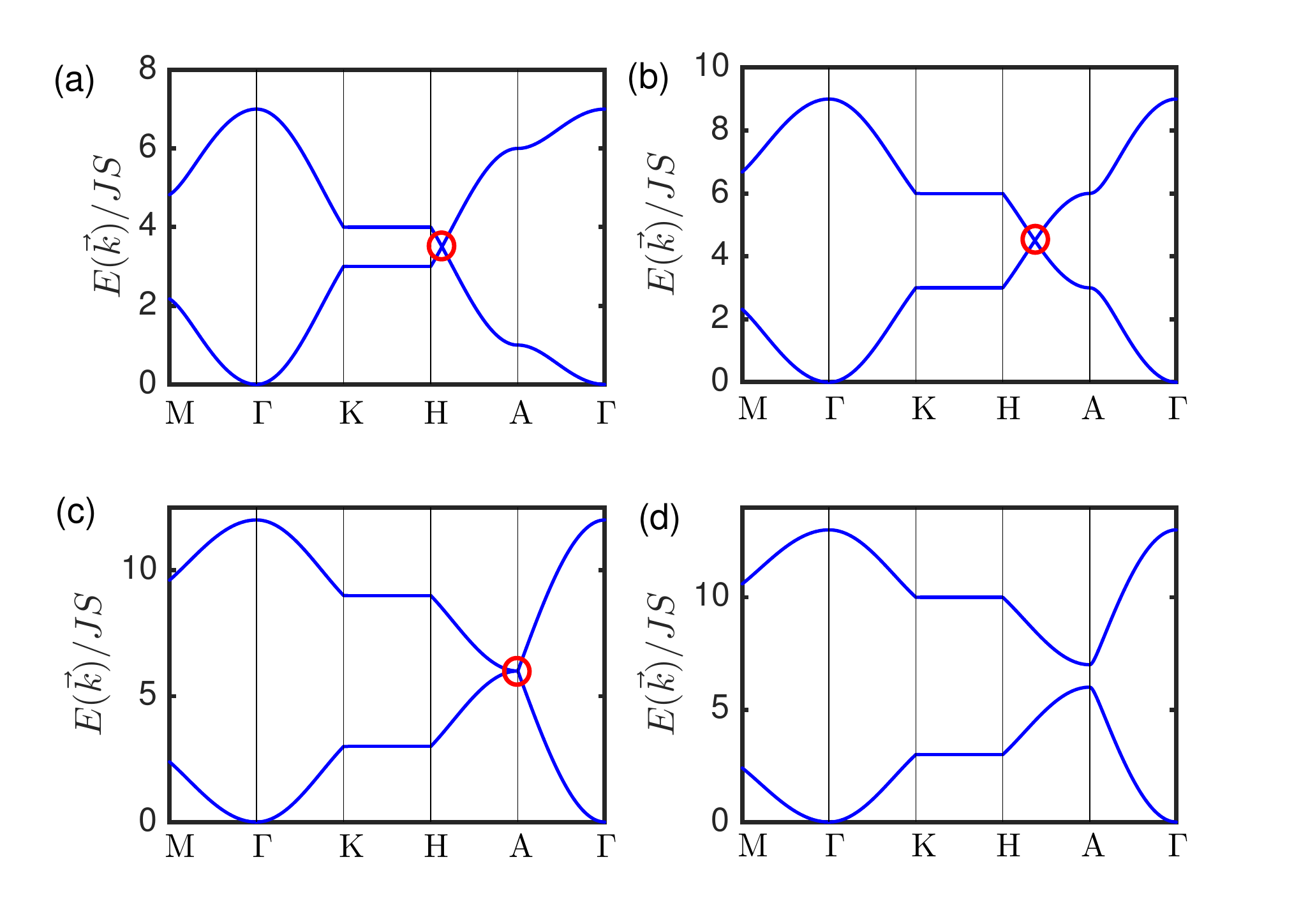}
\caption{Color online. Evolution of the magnon bands of undriven layered honeycomb ferromagnets, showing the phase transition from 3D DNL magnons to 3D gapped trivial magnon insulator. (a)  $t_L/t=0.5$, (b)  $t_L/t=1.5$, (c) $t_L/t=3$, (d) $t_L/t=3.5$. The red circles denote the DNLs.}
\label{udband}
\end{figure}

For  $t_L/t<3$, the DPs form  loops or rings centred at the $\bar{\text{K}}$-point in the (001) surface BZ, and thereby realize DNLs.  For the Dirac nodal loops centred at the $\bar{\text{K}}$-point in the $k_z=\pi$ plane,  the expression for the loops is $q_x^2+q_y^2=(t_L/v_s)^2$, where $v_s=3t/2$ is the group velocity, and $\vec q=\bar{\text{K}}-\vec k_\parallel$ is the momentum deviation from the DNL. In the regime $t_L/t>3$,  a gapped trivial insulator is obtained. In this report,  we will study both the DNLs and the gapped trivial insulator.   The phase transition  from  DNLs to gapped trivial insulator is depicted in Fig.~\eqref{udband}.  Note that in the vicinity of the DNLs at the $\bar{\text{K}}$-point, the functions $f_x(\vec k)$ and $f_y(\vec k)$ are linear in $k_x$ and $k_y$ respectively. Since $\mathcal T$-symmetry is preserved, the Berry curvature of the DNLs vanishes. Therefore, their topological protection is only characterized by the Berry phase defined as $\gamma=\oint_{\mathcal C} \mathcal A(\bo)\cdot d{ \bo} $, over a closed loop $\mathcal C$, where $\mathcal A(\bo)$ is the Berry connection given by $\mathcal A(\bo)=-i\braket{\psi_{\bo}^\dg|{\vec \nabla}_{\bo}\psi_{\bo}}$, and $\psi_{\bo}$ is the magnon eigenvectors. For a closed path encircling the DNLs in momentum space, the Berry phase is $\gamma=\pi$,  otherwise $\gamma=0$.

\vspace{10px}
\noindent\textbf{Driven magnonic Dirac nodal-line and Photoinduced Weyl magnons.}~~ The notion of periodically driven  magnonic systems  essentially rely on the quantum theory of magnon quasiparticles. Magnons are in fact moving magnetic dipoles in a magnetically ordered system and they are   charge-neutral bosonic quasiparticle with an intrinsic spin-$1$. The  magnetic dipole moment is given by ${\vec \mu}=\mu_m { \hat{z}}$,  where $\mu_m=g\mu_B$, $g$ is the spin $g$-factor and $\mu_B$ is the Bohr magneton. Therefore, magnons can couple to electric fields through the Aharonov-Casher effect \cite{aha,loss,ahaz,ahat}, similar to the coupling of charged particles through the Aharonov-Bohm effect \cite{aha1}. In general, a neutral particle couples non-minimally to an external electromagnetic field (see Methods). 

In the current study, we will apply this new concept to 3D insulating quantum ferromagnets possessing 3D DNL magnon and 3D gapped trivial magnon insulator phases.  We consider photo-irradiation of magnons in the insulating quantum ferromagnets described by the pristine Hamiltonian in Eq.~\eqref{model}. In the case of time-periodic electromagnetic field possessing a dominant time-dependent electric field components $\vec{E}(\tau)$, the effects of the electric field  can be described by a vector potential $\vec{ A}(\tau)$, where $\vec{ E}(\tau)=-\partial \vec{A}(\tau)/\partial \tau$.  The time-periodicity guarantees that  $\vec{A}(\tau+T)=\vec{A}(\tau)$, with $T=2\pi/\omega$ being the period.  In the real space geometry, this results in  a time-dependent Aharonov-Casher phase (see Methods)

\bea \theta_{\ell\ell^\prime}(\tau)=\mu_m\int_{\vec{r}_{\ell}}^{\vec{r}_{\ell^\prime}} \vec \Xi(\tau)\cdot d\vec{\ell},
 \label{eqn4}
 \eea 
 where $\mu_m = g\mu_B/\hbar c^2$ and $\vec{r}_\ell$ is the coordinate of the lattice at site $\ell$. We have used the notation $ \vec \Xi(\tau) = \vec{E}(\tau)\times \hat z$  for brevity.
 
  We will use the magnonic Floquet-Bloch theory develop in Methods, and consider specific form of the vector potential. A first choice would be a time-periodic  vector potential  in the $x$-$y$ plane given by $\vec{\Xi}(\tau)=[{E}_x\sin(\omega \tau), {E}_y\sin(\omega \tau+\phi),0]$ with amplitudes $E_x$ and $E_y$.  Here,  $\phi=\pi/2$ corresponds to circularly-polarized light  and  $\phi=0$ corresponds to linearly-polarized light. This form of vector potential is perpendicular to the DNLs and does not give any WPs \cite{we4a, we4}. In the magnonic honeycomb ferromagnetic system, the vector potential  in the $x$-$y$ plane gives rise to a photoinduced next-nearest-neighbour DM interaction in the $x$-$y$ plane pointing along the $z$-direction. This term breaks $\mathcal T$-symmetry, but yields a 3D magnonic Floquet TI  similar to the 2D system \cite{owe}.  Thus, there is no magnonic Floquet WPs for this choice of vector potential.

However, the 3D nature of the current model gives us another option for the vector potential. Now, we consider a different time-periodic vector potential in the $y$-$z$ plane given by $\vec{\Xi}(\tau)=[0, E_y\sin(\omega \tau),E_z\sin(\omega \tau+\phi)]$ with amplitudes $E_y$ and $E_z$.  This form of vector potential is parallel to the DNLs, hence  WPs are expected to emerge \cite{we4a, we4}. 

The time-dependent  Hamiltonian $\mathcal H({\vec k},\tau)$  is given by
\begin{align}
\mathcal H(\vec k,\tau)= \rho_0{\bf 1}_{2\times 2}+
\begin{pmatrix}
0&\rho(\vec k,\tau)\\
\rho^*(\vec k,\tau)&0\\
\end{pmatrix},
\label{tham1}
\end{align}
where $\rho(\vec k,\tau)=\rho(k_z,\tau)+\rho(\vec k_\parallel,\tau)$, $\rho(k_z,\tau) =-t_L e^{i(k_{z}+\mu_m\vec {\Xi}(\tau))}$ and $\rho(\vec k_\parallel,\tau) =-t \sum_j e^{i\lb\vec{k}_{\parallel}+ \mu_m\vec {\Xi}(\tau)\rb\cdot\vec{d}_j}$. The corresponding  Fourier components of the Hamiltonian \eqref{tham1} are given by

\begin{align}
\mathcal H_q(\vec k)&= \rho_0{\bf 1}_{2\times 2}+
\begin{pmatrix}
0&\rho_q(\vec k)\\
\rho_{-q}^*(\vec k)&0\\
\end{pmatrix}.
\label{fham1}
\end{align}
 \begin{figure}
\centering
\includegraphics[width=.8\linewidth]{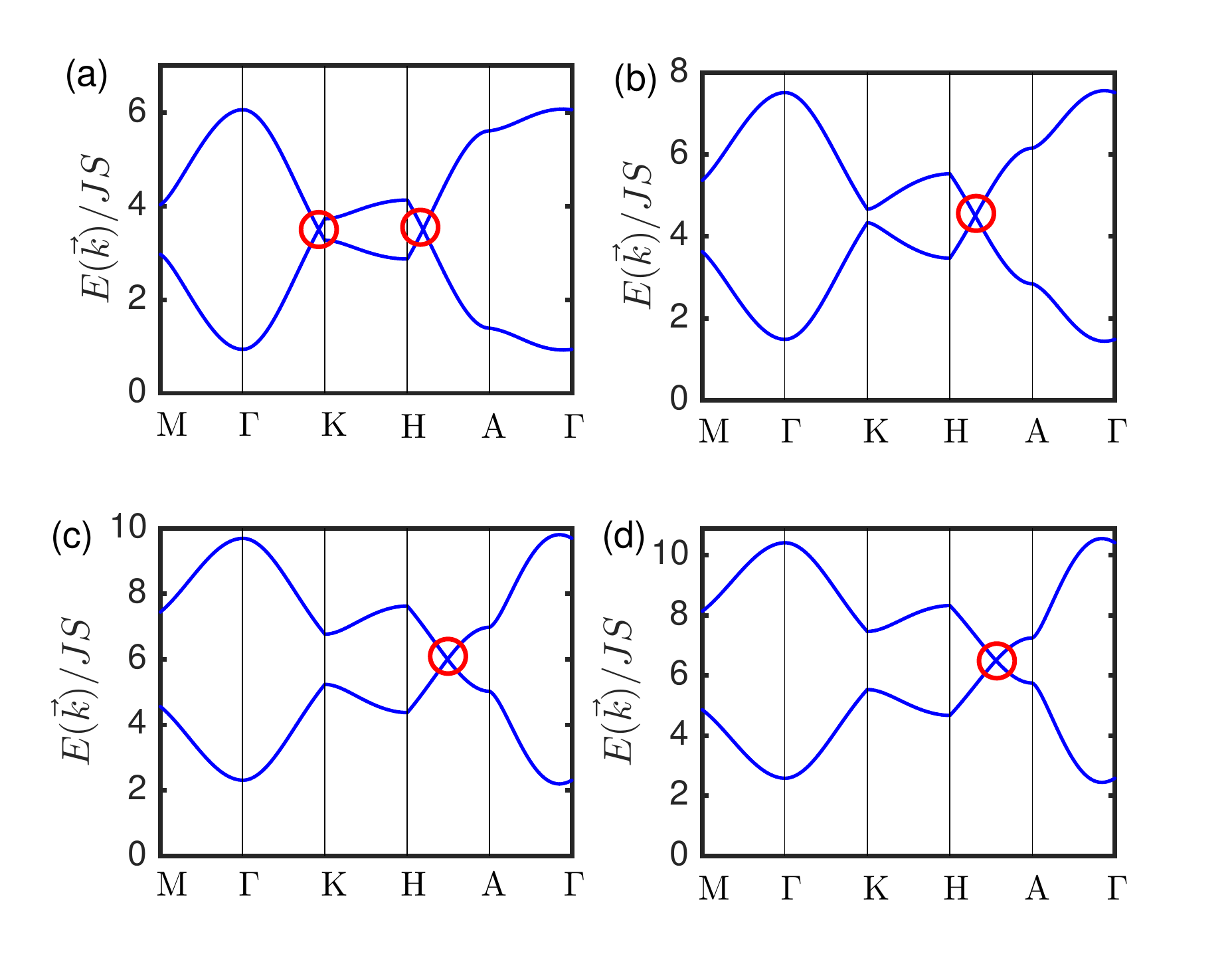}
\caption{Color online. Magnon bands of  periodically driven  layered honeycomb ferromagnets for  $\mathcal E_z=\mathcal E_y=1.7,~\phi=\pi/2$, and $\omega/t=10$. (a)  $t_L/t=0.5$, (b)  $t_L/t=1.5$, (c) $t_L/t=3$, (d) $t_L/t=3.5$. The  red circle denotes the photoinduced magnonic WPs.}
\label{F_band}
\end{figure}
 For the vector potential in the $y$-$z$ plane, we have 
 \begin{align}
 &\rho_q( k_z)=-t_L\mathcal J_q(\mathcal {E}_z)e^{ik_z}e^{iq\phi},\quad \rho_q(\vec k_{\parallel})=-\sum_{j=1}^3 t_{j, q} e^{i\vec{k}_{\parallel}\cdot\vec{d}_j},
 \end{align}
where the renormalized interactions in this case are given by $t_{1,q}=t\mathcal  J_{-q}(\mathcal {E}_y/2),~ t_{2,q}=t\mathcal J_{-q}(\mathcal {E}_y/2),~ t_{3,q}= t\mathcal J_{q}(\mathcal {E}_y)$. The dimensionless quantity characterizing the intensity of light is different from that of electronic systems and it is given by
\begin{align}
\mathcal{E}_i =\frac{g\mu_B E_i a}{\hbar c^2},
\end{align}
where $ \quad  i = x,y,z$ and $a$ is the lattice constant.

Next, we study the high frequency regime ($\omega\gg \Delta$), when the driving frequency $\omega$ is larger than the magnon bandwidth $\Delta$. In this regime the Floquet sidebands are decoupled, and the system can be described by a time-independent effective Hamiltonian \cite{tp6,tp7,tp8}, which can be obtained perturbatively in $1/\omega$ expansion as 
\begin{align}
\mathcal H_{\text {eff}}(\vec k)&=\mathcal H_0(\vec k)-\frac{1}{\omega}\big( \big[\mathcal H_0(\vec k), \mathcal H_{-1}(\vec k)\big]-\big[\mathcal H_0(\vec k), \mathcal H_{1}(\vec k)\big]+\big[\mathcal H_{-1}(\vec k), \mathcal H_{1}(\vec k)\big]\big),
\label{effHam}
\end{align}
where $\mathcal H_0(\vec k)$ is the zeroth order Hamiltonian and $\mathcal H_{\pm 1}(\vec k)$ are the single photon dressed Hamiltonians. In the effective model expanded near the $\bar{\text{K}}$-point, the first two commutators can be neglected. In the current system, however, we will not consider the  effective model near the crossing point, and thus there is no reason to neglect the first two commutators since they can have a nonzero contribution away from the $\bar{\text{K}}$-point.

\begin{figure}
\centering
\includegraphics[width=.8\linewidth]{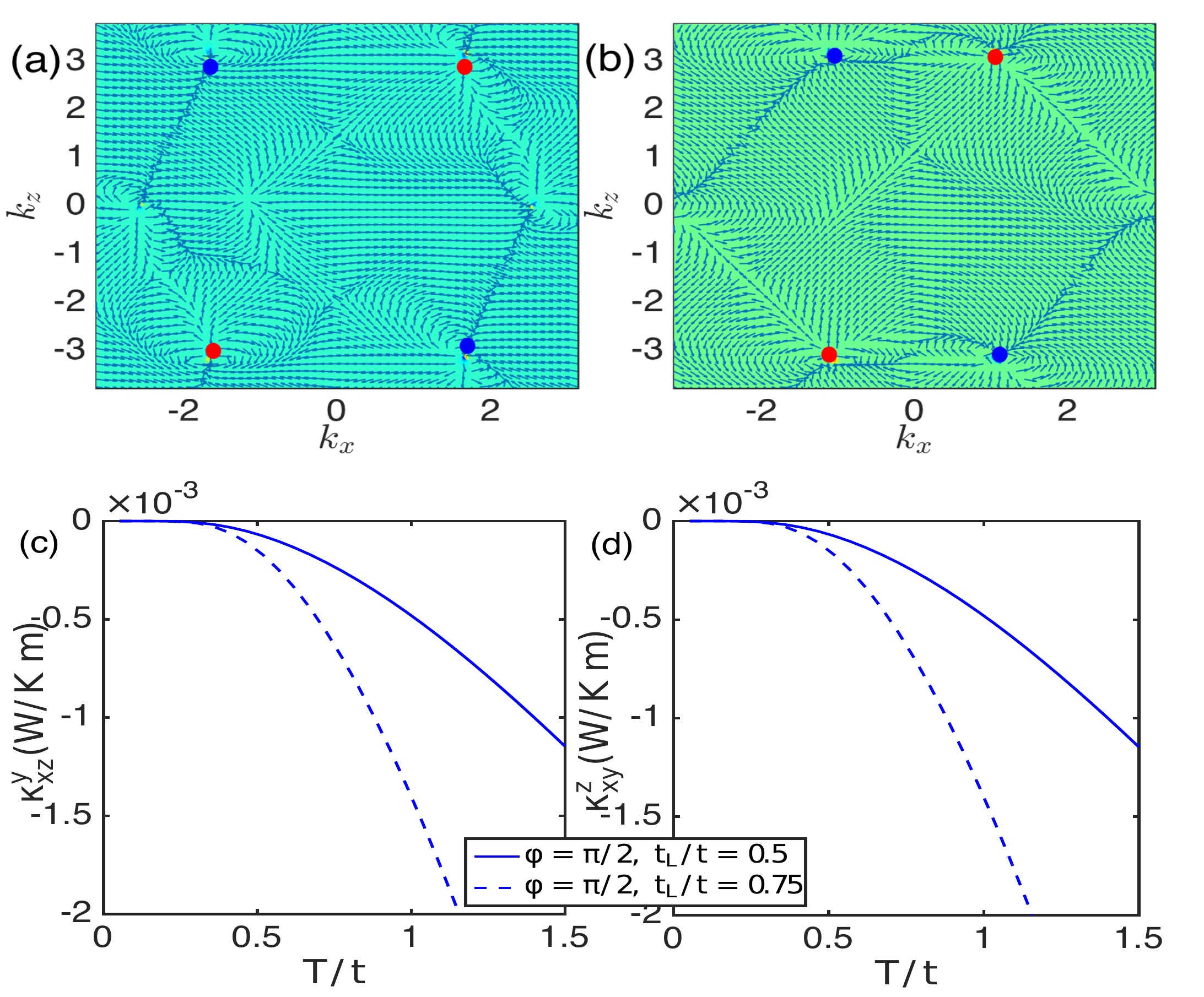}
\caption{Color online. Top panel. Monopole distributions  of the lowest magnon band Berry curvature $\Omega_{1,xz}^y(\vec k)$ for the photoinduced magnonic WPs at (a) $t_L/t=1.5$ and (b) $t_L/t=3.5$.  Bottom panel. The  thermal Hall conductivity in units of $k_B$. (c)~ $\kappa_{xz}^y$ vs. $T/t$ and  (d)~ $\kappa_{xy}^z$ vs. $T/t$. Here we set $\mathcal E_z=\mathcal E_y=1.7$, $\phi=\pi/2$, and $\omega/t=10$.}
\label{THE}
\end{figure}

 We have shown the effect of circularly-polarized light on the DNL magnons and the trivial magnon insulator in  Figs.~\ref{F_band}(a)--(c) and Fig.~\ref{F_band}(d) respectively.   For circularly-polarized light, {\it i.e.} $\phi=\pi/2$, we find that  the DNL magnons for $t/t_L<3$ are not gapped out, but transform to  photoinduced magnonic WPs as shown in Fig.~\ref{F_band}(a)--(c). Interestingly, circularly-polarized light also closes the gap in the  trivial magnon insulator phase for $t/t_L>3$, thereby generating  photoinduced magnonic WPs as shown in Fig.~\ref{F_band}(d). Thus, both rotational and time-reversal symmetries are broken by photo-irradiation. We note that additional linear magnon band crossings occur along $\Gamma$--$\rm{K}$ line depending on the model parameters. 
 
We have derived the expression for the effective Hamiltonian in Eq.~\eqref{effHam} (see Methods).  We find that the perturbative corrections to $\mathcal H_0(\vec k)$ gives a term proportional to $f_z(\vec k)\sigma_z$ in the effective Hamiltonian \eqref{effHam}.  Indeed, this term breaks $\mathcal{T}$-symmetry ({\it i.e.} $f_z(-\vec k)\neq f_z(\vec k)$), and thus imposes an additional  condition $f_x(\vec k)=f_y(\vec k)=f_z(\vec k)=0$ for magnon band crossing to occur.  The main result of this report is  that photo-irradiation in the direction  perpendicular to the DNLs generates 3D magnonic Floquet TIs, whereas photo-irradiation in the direction parallel to the DNLs  generates 3D magnonic Floquet WPs, in analogy to electronic systems \cite{we4a, we4}. We would also like to mention that the time-periodic vector potential in the $y$-$z$ plane does not generate a photo-induced next-nearest-neighbour DM interaction in the $x$-direction, since this term does not make any contribution to the magnon bands in linear spin wave theory.  Therefore, the  magnonic Floquet WPs in the current model do not originate from the out-of-plane DM interaction mechanism as opposed to magnonic WPs in the  undriven 3D quantum ferromagnets \cite{mw2,mw3,mw4,mw5}.



\vspace{10px}
\noindent\textbf{Monopoles of the Berry curvatures.}~~
The band structures of the undriven DNL magnons in Fig.~\eqref{udband}  are very similar to the corresponding photoinduced magnonic  WPs in Fig.~\eqref{F_band}. This suggests that the analysis of the magnon band structures  cannot sufficiently distinguish between DNLs and WPs. To  distinguish the two, we need to compute the Berry curvature associated with the magnon band  crossing points. As we noted above, the Berry curvature vanishes in the undriven DNLs as well as gapped trivial insulators as a result of  $\mathcal {T}$ symmetry. Therefore, a non-vanishing Berry curvature with linear magnon band crossing must be a consequence of WPs  due to broken  $\mathcal {T}$  symmetry. In general, WPs are the source or sink of the Berry curvature, which means that a single WP can be considered as a monopole of the Berry curvature in momentum space. 

We define the Berry curvature of a given magnon band $n$   as
\begin{align}
\Omega_{n,ij}^\ell(\vec k)=-2\text{Im}\sum_{n^\prime \neq n}\frac{[ \braket{\psi_{n}(\vec k)|\hat v_i|\psi_{ n^\prime}(\vec k)}\braket{\psi_{n^\prime}(\vec k)|\hat v_j|\psi_{n}(\vec k)}]}{[\epsilon_{n}(\vec k)- \epsilon_{n^\prime}(\vec k)]^2},
\label{chern2}
\end{align}
where $\hat v_{i}=\partial \mathcal{H}_{\text{eff}}(\vec k)/\partial k_{i}$ are the velocity operators,  $\psi_{n}(\vec k)$ are the magnon eigenvectors, and  $\epsilon_{ \alpha}(\vec k)$ are the  magnon quasi-energies. The  Berry curvature can be considered as a pseudo-vector pointing along  the $\ell$ directions perpendicular to both the $i$ and $j$ directions. All the components of the Berry curvature are found to be nonzero.  In the top panel of Fig.~\eqref{THE}, we have shown the plot of the monopole field  distributions of the lowest magnon band Berry curvature $\Omega_{\alpha,xz}^y(\vec k)$ (with $\alpha=1$) in the $k_y=0$ plane. We note that the Berry curvature is maximized at the photoinduced magnonic WPs.  The monopole  distribution of the Berry curvature is a compelling evidence that the photoinduced magnon band crossings  are indeed magnonic WPs.

\vspace{10px}
\noindent\textbf{Anomalous thermal Hall effect.}~~
 In analogy to anomalous Hall effect in electronic Weyl semimetals \cite{abur, kyang}, the  magnonic WPs in 3D quantum magnets also exhibit  the anomalous thermal Hall effect \cite{mw7}, which is generally understood as a consequence of the nonvanshing Berry curvatures. In the  high frequency limit,  the system is close to equilibrium. Thus, the same theoretical concept of undriven anomalous thermal Hall effect due to a temperature gradient \cite{kat,alex2} is applicable to the driven system close to thermal equilibrium.  The transverse components $\kappa_{ij}^{\ell}$ of the thermal Hall conductivity are given by \cite{alex2, alex2q}
\begin{align}
\kappa_{ij}^{\ell}=-k_B^2 T\int_{BZ} \frac{d^3k}{(2\pi)^3}~ \sum_{n=1}^N c_2\lb f_n^B\rb\Omega_{n,ij}^{\ell}(\vec k).
\label{thm}
\end{align}

Here, $ f_n^B=1/ \big(e^{\epsilon_{n}(\vec k)/k_BT}-1\big)$ is the Bose distribution function close to thermal equilibrium,  $k_B$ is the Boltzmann constant, $T$ is the temperature, and $ c_2(x)=(1+x)\lb \ln \frac{1+x}{x}\rb^2-(\ln x)^2-2\text{Li}_2(-x)$, with $\text{Li}_2(x)$ being the  dilogarithm. Similar to the Berry curvature, $\kappa_{ij}^{\ell}$ vanishes in the undriven DNLs and gapped trivial insulators due to $\mathcal T$-symmetry. The maximum contribution to $\kappa_{ij}^{\ell}$  comes from the photoinduced magnonic WPs at the lowest magnon excitation due to the Berry curvature. It can be shown that $\kappa_{ij}^{\ell}$ depends on the distribution of magnonic WPs  in momentum space  \cite{mw7}, in analogy  to the thermal Hall effect in electronic Weyl semimetals \cite{ferre}. In the bottom panel of Fig.~\eqref{THE}, we have shown the  trends of  (c)~ $\kappa_{xz}^y$ and  (d)~$\kappa_{xy}^z$ in the photoinduced Weyl magnon phase.

\vspace{10px}
\noindent\textbf{Conclusion}

The main result of this report is that magnonic WPs can be photoinduced in three-dimensional (3D)  quantum magnets  initially possessing   DNL magnon and gapped trivial magnon insulator phases. We  achieved this result by utilizing magnons as hopping magnetic dipole moment in an ordered quantum magnet. Hence, magnons couple to time-dependent electric field through the time-dependent Aharonov-Casher effect as shown in Methods. In other words, the electric charge in electronic systems is dual to the magnetic dipole moment in magnonic systems.   The newly proposed magnonic Floquet WPs have many advantages over intrinsic magnonic WPs. First, they can be tuned by the incident light, and can also be engineered in different magnetic systems. Second, they do not rely on intrinsic perturbative interactions to break time-reversal symmetry, and they could also provide a platform for investigating  new features such as  photo-magnonics \cite{benj}, magnon spintronics \cite{magn, benja},  and ultrafast optical control of magnetic spin currents \cite{ment, tak4, tak4a,walo}. Therefore, the current results  are also pertinent to experimental investigation, and can be applied to different bulk 3D quantum magnetic materials. Thereby, extending the notion of  magnonic WPs to a broader class of 3D quantum magnets. 

We note that there is very little spectroscopic experimental progress in the observation of magnonic analogs of electronic topological systems. Recently, bulk Dirac magnons have been experimentally confirmed  in the 3D antiferromagnet Cu$_3$TeO$_6$ \cite{kli, yao,bao}. The measurement of the anomalous thermal Hall effect \cite{alex1, alex1a, alex6} is also an alternative way to confirm the existence of topological spin excitations in quantum magnets. It should be noted that the thermal Hall effect is absent in the undriven Dirac and nodal-line magnons, as well as the trivial magnon insulators, because of the presence of time-reversal-symmetry.  Moreover, the chiral magnon edge and surface magnon modes are yet to be verified experimentally in topological magnon systems. In this report,  we have focused on features that can be directly measured experimentally e.g. by using  ultrafast terahertz spectroscopy, inelastic neutron scattering, and thermal Hall measurements.

\vspace{10px}
\noindent\textbf{\large Methods} 

\noindent\textbf{Magnonic Floquet-Bloch theory.}~~ The Floquet-Bloch theory is a formalism for studying  periodically driven quantum systems and it applies to different cases of physical interests. The magnonic version describes the interaction of light with magnonic Bloch states in insulating quantum magnets.  In the present case, the  time-dependent Hamiltonian $\mathcal H(\vec{k},\tau)$ can be obtained by making the  time-dependent Peierls substitution  ${\vec k}\to {\vec k} +\vec{\mathcal A}(\tau)$. Note that $\mathcal H(\vec{k},\tau)$   is periodic due to the time-periodicity of the vector potential. 

Hence, it can be expanded in Fourier space as
$\mathcal{H}(\vec{k},\tau)= \mathcal{H}(\vec{k}, \tau+T)=\sum_{n=-\infty}^{\infty} e^{in\omega \tau}\mathcal{H}_n(\vec{k}),$ where $\mathcal{H}_n(\vec{k})=\frac{1}{T}\int_{0}^T e^{-in\omega \tau}\mathcal{H}(\vec{k}, \tau) d\tau=\mathcal{H}_{-n}^\dg(\vec{k})$ is the Fourier component. Thus, we can write its eigenvectors in the Floquet-Bloch form $\ket{\psi_\alpha(\vec{k}, \tau)}=e^{-i \epsilon_\alpha(\vec{k}) \tau}\ket{\xi_\alpha(\vec{k}, \tau)}$, where  $\ket{\xi_\alpha(\vec{k}, \tau)}=\ket{\xi_\alpha(\vec{k}, \tau+T)}=\sum_{n} e^{in\omega \tau}\ket{\xi_{\alpha}^n(\vec{k})}$ is the time-periodic Floquet-Bloch wave function of magnons and $\epsilon_\alpha(\vec{k})$ are the magnon quasi-energies. We define the Floquet operator as $\mathcal{H}^F(\vec{k},\tau)=\mathcal{H}(\vec{k},\tau)-i\partial_\tau$, which  leads  to the Floquet eigenvalue equation
\begin{align}
\sum_m [\mathcal H^{n-m}(\vec{ k}) + m\omega \delta_{n,m}]\xi_{\alpha}^m(\vec{k})= \epsilon_\alpha(\vec{k})\xi_{\alpha}^n(\vec{k}).
\end{align}
 \vspace{10px}

\noindent\textbf{Massless neutral particle  in an external electromagnetic field.}~~
In this section, we give the general theory for a massless neutral particle with magnetic dipole moment such as the magnonic DNL quasiparticle,  coupled non-minimally to an external electromagnetic field  (denoted by the tensor $ F_{\mu\nu}$) via its magnetic dipole moment ($\mu$). In (3+1) dimensions, the system is described  by the Dirac-Pauli Lagrangian \cite{bjo} 
\begin{align}
\mathcal L=\bar\psi(x)(i\gamma^\mu\partial_\mu-\frac{\mu}{2}\sigma^{\mu\nu} F_{\mu\nu})\psi(x),
\end{align}
where $\hbar=c=1$ has been used. Here $x\equiv x^\mu=(x^0,\vec x)$, $\bar\psi(x)=\psi^\dg(x)\gamma^0$, and $\gamma^\mu=(\gamma^0,\vec\gamma)$ are the $4\times 4$  Dirac matrices that obey the algebra \bea \lbrace \gamma^\mu,\gamma^\nu\rbrace=2g^{\mu\nu},~\text{where}~ g^{\mu\nu}=\text{diag}(1,-1,-1,-1),\eea and  \bea\sigma^{\mu\nu}=\frac{i}{2}[\gamma^\mu,\gamma^\nu]=i\gamma^\mu\gamma^\nu,\quad (\mu\neq \nu).\eea 

In this report, we will consider the system with  only spatially uniform and time-varying electric field  $\vec{\mathcal E}(\tau)$. In this case, the corresponding Hamiltonian is given by
\begin{align}
\mathcal H=\int d^3 x ~\psi^\dg(x)\big[\vec{\alpha}\cdot\big(-i\vec{\nabla}-i\mu\beta\vec{\mathcal E}(\tau)\big)\big]\psi(x),
\end{align}
where $\vec{\alpha}=\gamma^0\vec \gamma$ and $\beta=\gamma^0$. 

In (2+1) dimensions,   the Dirac matrices  are simply Pauli matrices given by
\begin{align}
\beta=\gamma^0=\sigma_z,~\gamma^1=i\sigma_y,~\gamma^2=-i\sigma_x.
\end{align}
The corresponding momentum space Hamiltonian in (2+1) dimensions now takes the form
\begin{align}
\mathcal H=\int \frac{d^2k}{(2\pi)^2}~\psi^\dg(\bo,\tau)\mathcal H(\bo,\tau)\psi(\bo,\tau),
\end{align}
where
\begin{align}
\mathcal H(\bo,\tau)=\vec{\sigma}\cdot\big[\bo+\mu_m\big(\vec{E}(\tau)\times \hat z\big)\big],~\text{with}~\vec{\sigma}=(\sigma_x,\sigma_y).
\label{nl1}
\end{align}

 We see that the Hamiltonian in Eq.~\eqref{nl1} is equivalent  to that of DNL Hamiltonian in Eq.~\eqref{nl} near  the crossing point, coupled to a time-periodic electric field through the  magnetic dipole moment $\vec \mu_m=\mu_m\hat z$.  
The time-dependent Aharonov-Casher phase is evident from the Hamiltonian in Eq.~\eqref{nl1}. 

\vspace{10px}
\noindent\textbf{Effective Hamiltonian.}~~
In this section, we derive the form of the effective Hamiltonian in Eq.~\eqref{effHam} in the case of a vector potential in the $y$-$z$ plane. The effective Hamiltonian can be written as 
\begin{align}
\mathcal H_{{\rm eff}}(\vec k)=f_0\sigma_0 +f_x^0(\vec k)\sigma_x+f_y^0(\vec k)\sigma_y+f_z(\vec k)\sigma_z,
\label{efnl}
\end{align}
where,
\begin{align}
&f_x^0(\vec k)=-\sum_{j=1}^3 t_j^0\cos(\vec k_\parallel\cdot \vec d_j)-t_L^0\cos(k_z),\quad 
f_y^0(\vec k)=\sum_{j=1}^3 t_j^0\sin(\vec k_\parallel\cdot \vec d_j)+t_L^0\sin(k_z),
\end{align}
with  $t_{1}^0=t\mathcal  J_{0}(\mathcal {E}_y/2),~ t_{2}^0=t\mathcal J_{0}(\mathcal {E}_y/2),~ t_{3}^0= t\mathcal J_{0}(\mathcal {E}_y)$, and $t_L^0=t_L\mathcal J_{0}(\mathcal {E}_z)$.
\begin{align}
f_z(\vec k)&=\frac{4}{\omega}\Bigg[2t^2\mathcal  J_{0}(\mathcal {E}_y/2)\mathcal  J_{1}(\mathcal {E}_y/2)+t^2\mathcal  J_{0}(\mathcal {E}_y)\mathcal  J_{1}(\mathcal {E}_y)+2tt_L\mathcal{J}_{0}(\mathcal {E}_z)\mathcal  J_{1}(\mathcal {E}_y)\cos(\sqrt{3}k_x) \nonumber\\& + tt_L\mathcal{J}_{0}(\mathcal {E}_z)  \mathcal{J}_{1}(\mathcal {E}_y)\cos(k_y-k_z)+\mathcal{J}_{1}(\mathcal {E}_z) \lbrace t_L^2 \mathcal{J}_{0}(\mathcal {E}_z)+tt_L \mathcal{J}_{0}(\mathcal {E}_y)\cos(k_y-k_z)\rbrace\cos(\phi)\nonumber\\& + 2\cos(\sqrt{3}k_x/2)\big[ t^2\lbrace  \mathcal{J}_{0}(\mathcal {E}_y) \mathcal{J}_{1}(\mathcal {E}_y/2)+\mathcal{J}_{0}(\mathcal {E}_y/2) \mathcal{J}_{1}(\mathcal {E}_y) \rbrace\cos(\sqrt{3}k_y/2)\nonumber\\&  +tt_L\lbrace \mathcal{J}_{0}(\mathcal {E}_z) \mathcal{J}_{1}(\mathcal {E}_y/2)+\mathcal{J}_{0}(\mathcal {E}_y/2) \mathcal{J}_{1}(\mathcal {E}_z)\cos(\phi) \rbrace\cos(\sqrt{3}k_y/2+k_z)\big] \nonumber\\&-tt_L\mathcal{J}_{1}(\mathcal {E}_z)\sin(\phi)\lbrace \mathcal{J}_{1}(\mathcal {E}_y)\sin(k_y-k_z)-2\mathcal{J}_{1}(\mathcal {E}_y/2)\cos(\sqrt{3}k_x/2)\sin(k_y/2+k_z)\rbrace\Bigg].
\end{align}

\vspace{10px}
\noindent\textbf{Acknowledgements}

\noindent Research at Perimeter Institute is supported by the Government of Canada through Industry Canada and by the Province of Ontario through the Ministry of Research
and Innovation. 

\vspace{10px}
\noindent\textbf{Author Contributions}

\noindent S. A. Owerre conceived the idea, performed the calculations, discussed the results, and wrote the manuscript.

\vspace{10px}
\noindent\textbf{Additional Information}

\noindent\textbf{\small Competing Interests}. I declare that the author has no competing interests as defined by Nature Research, or other interests that might be perceived to influence the results and/or discussion reported in this paper.


\begin{thebibliography}{99}
\bibitem{foot}
In contrast to undriven magnonic WPs  induced by the  $z$-component of the DM vector \cite{mw2,mw3,mw4,mw5}, we will show  that the $z$-component of the  photoinduced DM interaction is not the source of the magnonic Floquet WPs in the current 3D ferromagnetic system.
\end{thebibliography}
\end{document}